\documentclass[11pt,twoside]{article}
\usepackage{amsfonts}
\usepackage{amsmath,amsthm,latexsym,amssymb}
\usepackage[latin1]{inputenc}
\usepackage{ifpdf}
\usepackage{float}
\usepackage{color}
\usepackage{graphics}
\usepackage[dvips]{graphicx}

\numberwithin{equation}{section}
\usepackage{caption}
\usepackage{subcaption}
\usepackage{wrapfig}

\begin{document}



\title {Exploring the roles of local mobility patterns, socioeconomic conditions, and lockdown policies in shaping the patterns of COVID-19 spread}


\author{Mauricio Herrera\thanks{Facultad de Ingenier\'{i}a, Universidad del Desarrollo (mherrera@ingenieros.udd.cl)}
	 \and Alex Godoy-Faúndez\thanks{ Facultad de Ingenier\'{i}a, Universidad del Desarrollo.
Av. Plaza 680, Las Condes, Santiago de Chile, Chile. (alexgodoy@ingenieros.udd.cl)}  }


\date{\today}
\date{}
\maketitle

\begin{abstract}
The COVID-19 crisis has shown that we can only prevent the risk of mass contagion through timely, large-scale, coordinated, and decisive actions. However, frequently the models used by experts [from whom decision-makers get their main advice] focus on a single perspective [for example, the epidemiological one] and do not consider many of the multiple forces that affect the COVID-19 outbreak patterns.  The epidemiological, socioeconomic, and human mobility context of COVID-19 can be considered as a complex adaptive system. So, these interventions (for example, lock-downs) could have many and/or unexpected ramifications. This situation makes it difficult to understand the overall effect produced by any public policy measure and, therefore, to assess its real effectiveness and convenience.
By using mobile phone data, socioeconomic data, and COVID-19 cases data recorded throughout the pandemic development, we aim to understand and explain [make sense of] the observed heterogeneous regional patterns of contagion across time and space. We will also consider the causal effects produced by confinement policies by developing data-based models to explore, simulate, and estimate these policies' effectiveness. We intend to develop a methodology to assess and improve public policies' effectiveness associated with the fight against the pandemic, emphasizing its convenience, the precise time of its application, and extension. The contributions of this work can be used regardless of the region. The only likely impediment is the availability of the appropriate data. 
\end{abstract}

\section{Introduction}

The COVID-19 pandemic has highlighted the critical importance of generating rigorous evidence for decision-making, and actionable insights from all the available data. In this work, by using recorded human mobility data and socioeconomic data, such as income, wealth, public health infrastructure, cultural practices and so on, we aim to understand/explain the observed regional patterns in the evolution of the pandemic. Human mobility plays an important role in the spread of contagions and its study helps to understand the patterns of COVID-19 outbreaks and their subsequent spread\footnote{The SARS of 2003, the H1N1 pandemic of 2009, the West African Ebola epidemic of 2014 provide some previous examples that confirm the importance of human mobility patterns on the speed and extent of the spread of infectious disease}. Mobile sensing data, on the other hand, can be used to measure/describe human mobility. This work is closely linked to a rapidly growing literature using mobile phone geolocation and other mobile sensing data to assess the spread of COVID-19. $Grandata$ (our data provider) have aggregated information of a sizable sample of smart phone users locations across twelve Latin American countries. In this paper, we use part of this information, registered for the Metropolitan Region of Santiago de Chile (MR), to describe/analyze mobility patterns in urban-sectors and its influences on COVID-19 spread. 
We complement these mobility data with (1) counts of coronavirus cases at the communal level from the COVID-19 data repository of the Ministry of Sciences, (2) mobility data from the origin-destination matrices obtained from the metropolitan public transport, (3) mobility indices calculated from the cell phone tower network,  and (4) publicly available socioeconomic on MR communes that include income, education, health-care access, cultural practices, and so on. 

This work also highlights the value of public policies related to mobility restrictions and its heterogeneous effectiveness to curb the spread in a developing country like Chile. In fact, Latin American countries are unable to adopt some of the measures that high-income countries have implemented due to less availability of resources and less preparedness of the countries \cite{who1}. Therefore, it may be interesting to learn details about the different approaches that developing countries have implemented to deal with the pandemic. Latin American countries have managed to control the pandemic through different strategies. In the cases of Argentina, Bolivia, Colombia and Peru, for example, total quarantine was established when the number of confirmed cases was still low. Some of these countries have been more successful compared to others, but most of them have suffered long periods of restricted mobility, showing clear signs of fatigue and successive outbreaks of infection. 

The Chilean government's strategy in the initial phase of the pandemic was to close educational establishments, close all non-essential businesses and declare a state of emergency with a night curfew. This partially restricted people's mobility for several specific hours during the night. Furthermore, in the capital Santiago, the government used relatively massive testing and implemented a dynamic quarantine strategy by dividing into sectors delimited by communes\footnote{Communes are administrative boundaries or fundamental administrative units in Chile.}, and further dividing and isolating specific geographic locations of those communes. These lockdowns were periodically reassessed and lifted, prolonged or expanded depending on the active cases in the sectors. After this first phase characterized by dynamic quarantines, the government declared a prolonged quarantine for practically the entire MR as of May 15. From this quarantine, the communes gradually left. The third phase of this period of the pandemic was given by the implementation of the plan called ``Step by Step''\cite{paso-paso}. Measures to restrict mobility in Chile are currently governed by this plan.

This study could contribute to the analysis of the management that different countries have made of the crisis and its very dissimilar results. In particular, in the case of Chile and despite the fact that the government used the same tools - early border closures, confinement policies, and massive tests - that other countries with more success in controlling the pandemic, such as New Zealand or Vietnam,  the result of its handling was considered among the ten worst in the world \cite{lowy}. Could we ask ourselves,  what are the reasons for this?.

This work also aims to provide inference procedures for the effects of confinement public policies through the construction of data-driven models, incorporating different drivers and fitting the parameters of the model using data already registered, in order to describe the course of the pandemic. These models seek to explain the effects of confinement policies on mobility patterns in a coherent way and their effects on the evolution of the pandemic. Effects that were also nuanced by comunnes socioeconomic conditions.

To provide some context, Santiago MR holds 8.125.072 people, accounting for approximately 42\% of the country's population, and has high population density of 527.5 people per km$^2$. This makes the MR the main focus of confirmed cases in the country, as of  October 29 accounting for 57\% of the COVID-19 cases\footnote{In the first months of the pandemic this percentage was even much higher, for example, as of May 15, it was 74\%} \cite{minsalud1}. 

Using observable variables obtained directly from the data, for example, (1) the number of trips recorded by locations of successive cell towers (Base-Transceivers-Stations) that cover the user's cell phone service during trips in the sector studied, (2) the displacement events measured in each hexagon \cite{hex} covering the studied sector\footnote{We mean, H3 geospatial indexation system. See section \ref{mobpattern}}, we characterize the observed mobility patterns in the MR\footnote{In principle, the models, methodology, and analyses in this work can be used regardless of the region. The only likely impediment is the availability of similar data}. We also consider (3) commune's socioeconomic conditions by using census data, incorporating this way, several economic and demographic indicators into the analysis.


 We build relationships (regressions) between these variables and analyze causal inferences. The number of COVID-19 cases is treated with these regression models along with synthetic and counterfactual control models directly applied to time series.

We find correlations between mobility patterns, along with socioeconomic drivers with the observed spread patterns of COVID-19. We also explore the impact produced by successive policies of confinement. The main goal is to  make a consistent narrative, scientifically validated, and data-supported to explain MR patterns of the pandemic using data on mobility, socioeconomic conditions, and public policies of confinement.

The proposed methodology and the tools used in this work are not exclusive to the case of Chile and would eventually also contribute to answering several related questions, for example: What is the precise moment to apply the confinement policy, to what extent and when to lift these restricted mobility measures? What are the possible results of not applying the mobility restrictions? How could these modifications affect the spread of the virus? How do socioeconomic conditions influence this spread?, etc.

\section{The datasets.} \label{datasets}

The first source of data used in this work, is that collected by Grandata\footnote{This dataset is used in the context of the UNDP LAC call "Exploring the impact and response to the COVID-19 pandemic in Latin America and the Caribbean using mobility data" sponsored (no funding, data only) by the Research Team/Policy Response Office of the Chief Economist, Regional Office for Latin America and the Caribbean.} \cite{grandata}. Since march 1, 2020, geolocation events of smartphone owners were recorded by the Grandata group, using a MAID (Mobile Advertising ID) ``hash''. These events track the mobility patterns of smartphone users for a considerable sample of the population of twelve Latin American countries. Geo-referenced data is presented for hexagonal areas constructed using the H3 geospatial indexing system \cite{hex}. Mobility data is aggregated at different resolution levels. We used level 6 hexagons in the H3 geospatial system, which spans 0.74 $km^2$.

\begin{figure}[H]
	\centering
	\includegraphics[trim = 0cm 0cm 0cm 0cm, width = 1.0\textwidth]{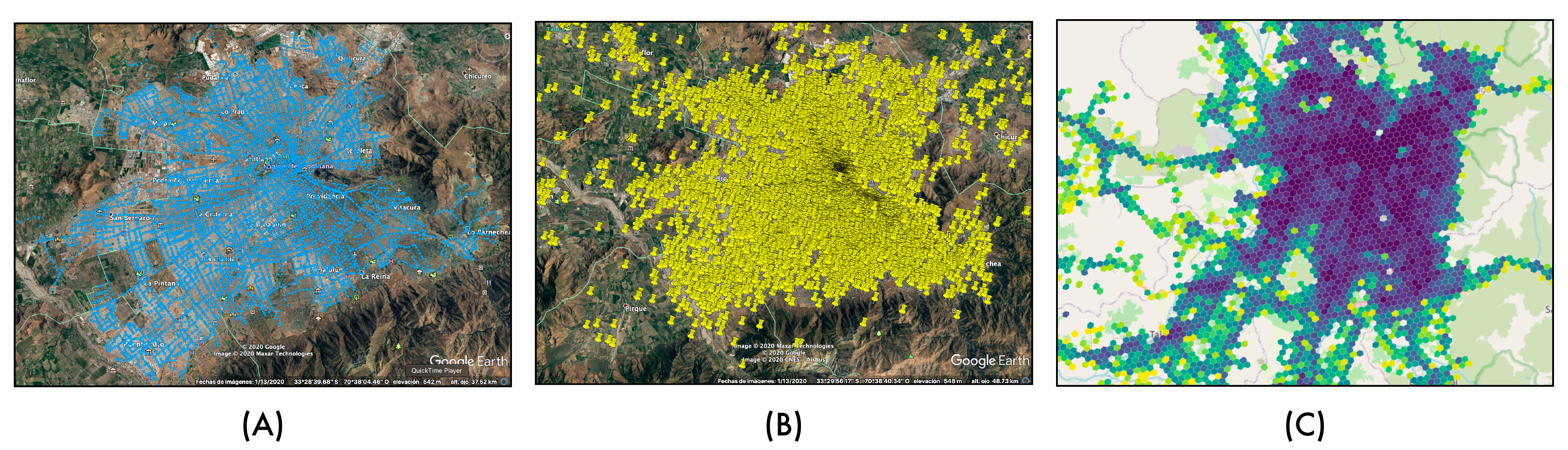}
	\caption{ Towards the capture of mobility patterns in MR through three sensing methodologies. Panel (A) shows around 12.000 bus stations in the MR public transport network. The OD matrices are estimated by the flow of passengers between these bus stops. Panel (B) indicates the network of cell phone antennas that provide coverage in the MR. The mobility index is determined by the flow of mobile phone users and the records that they leave on the antennas that cover the service while they are moving. Panel (C) shows level 6 hexagons covering the MR on a given day. The highest scores/percentile are in blue, while the lowest are yellow, passing through a range of greens.
	}
	\label{fig:1}
\end{figure}

The MR is covered with hexagons, in each of them a $Score$ is determined from the displacements that occur inside. The values of these scores indicate daily percentiles of the movements/mobility that take place within the area marked by the hexagon. The percentiles are calculated with respect to the distribution of all the displacements given by all the hexagons that cover the MR. In panel C of figure \ref{fig:1}, these scores are plotted for a given day. The highest scores/percentiles are in blue, while the lowest are yellow, going through a range of greens. 

 For MR, in Santiago de Chile, we use two other sources of mobility data. The first one is the origin-destination (OD) matrices \cite{bip}. Mobility data from the OD matrices are registered using ADATRAP software developed by Universidad de Chile and the Metropolitan Public Transport Directory \cite{bip}. This software obtains the OD matrices for trips on public transport (buses and Metro) by registering the use of payment cards and GPS of the buses  during the day and every half hour. The pick-up and drop-off locations of the trips are registered. Around 12.000 bus stations are considered (see figure \ref{fig:1} panel A). With these data, for example, trips between communes can be registered (see figure \ref{fig:11}). 
 
  The second source of data are the mobility indices built from the geolocation registered by the cellular telephone antenna network (see figure \ref{fig:1} panel B) of the Telefónica company \cite{movistar}.  Two indices are calculated, the internal mobility index (MobIn), which measures only trips within a commune. That is, the changes in the antennas and their respective locations, which give coverage to the cell phone service when the user moves within the limits of the commune. The external mobility index (MobOut), for its part, estimates the displacements of users outside the commune, taking into account the change of antennas that provide coverage to the different communes through which the user travels. The sum of these indices is called the \textit{Mobility Index}. 

 To assess socioeconomic indicators of Santiago Metropolitan communes we use census data and a more ad hoc, publicly available, data - the Territorial Well-being Index of 2012 \cite{uai}, which indexes the mean income of every census administrative unit down to the block level, of which Santiago has 39901. Another source of data is a publicly available COVID-19 cases data, registered on daily basis by the Ministry of Sciences \cite{minciencia}.

\section{Santiago Metropolitan Region mobility patterns.} \label{mobpattern}

Grandata mobility data contains logs of geolocation events tracking mobility patterns of the smartphone by using Mobile Advertising ID hash \cite{grandata}. Mobility data was constructed for hexagon shapes using the H3 geospatial system \cite{hex}. The geolocation events (exits/entrances to the hexagon) were summed for each area covered by hexagons, without considering the area of residence of the people. The ``Score'' for the each hexagon area consists of hexagon's percentile placement in the mobility distribution for a particular date, when taking into account all other hexagons in the corresponding administrative area (Santiago Metropolitan Area). The figure \ref{fig:1}(C) shows the scores or percentiles for hexagons covering the MR.

\begin{figure}[H]
	\centering
	\includegraphics[trim = 0cm 0cm 0cm 0cm, width = 0.8\textwidth]{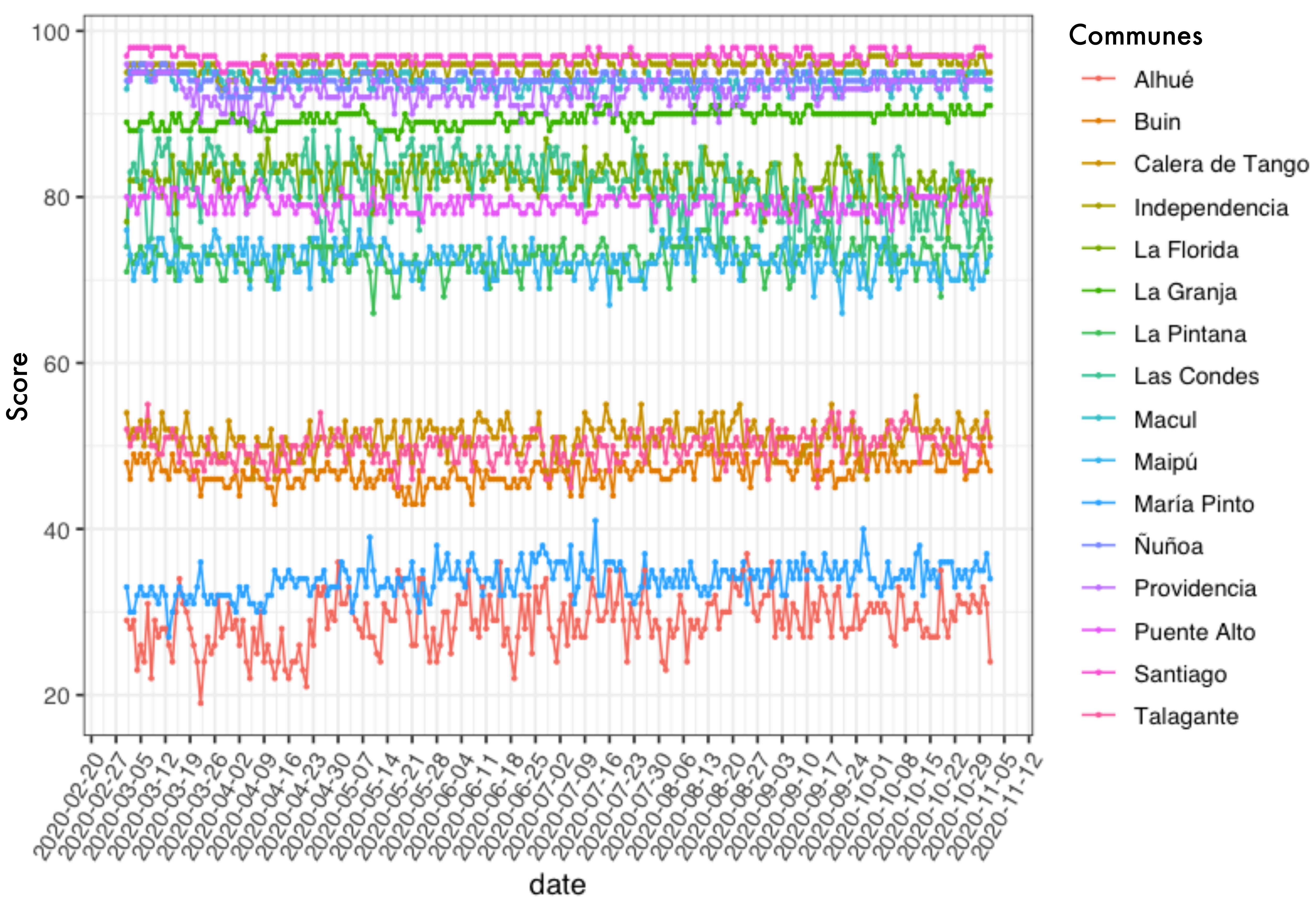}
	\caption{ Time series for the average mobility Scores of a sample of 16 RM communes.}
	\label{fig:6}
\end{figure}

In this section, we are going to study the mobility patterns of the MR considering the communes as the basic geographic units/sectors in our analysis. The MR is made up of 52 communes. We will assume that each H3 hexagon is a kind of ``sensor'' to measure the mobility of some commune area. Each commune is covered by a specific number of sensors-hexagons (the MR is covered by a total of 11.853 hexagons) that measure the mobility of the area for 249 days (time interval between 2020-03-02 and 2020-11-01), thus generating time series with mobility scores. We average the daily Scores of all the sensor-hexagons that make up each commune, as a proxy for the mobility of the communes during the day. In this way, we obtain 52 time series corresponding to the average Scores of each commune during the study time interval. The figure \ref{fig:6} shows these Scores time series for a sample of 16 MR communes.

\subsection{Time series clustering analysis.}

To capture the mobility patterns we use cluster analysis for time series. We use Dynamic Time Warping (DTW). DTW is a popular technique for comparing time series, providing both a distance measure that is insensitive to local compression and stretches and the warping which optimally deforms one of the two input series onto the other. The rationale behind DTW is, given two time series, to stretch or compress them locally in order to make one resemble the other as much as possible \cite{dtw}. For this analysis we used the R package \textbf{dtw} \cite{dtwR}.

\begin{figure}[H]
	\centering
	\includegraphics[trim = 0cm 0cm 0cm 0cm, width = 1.0\textwidth]{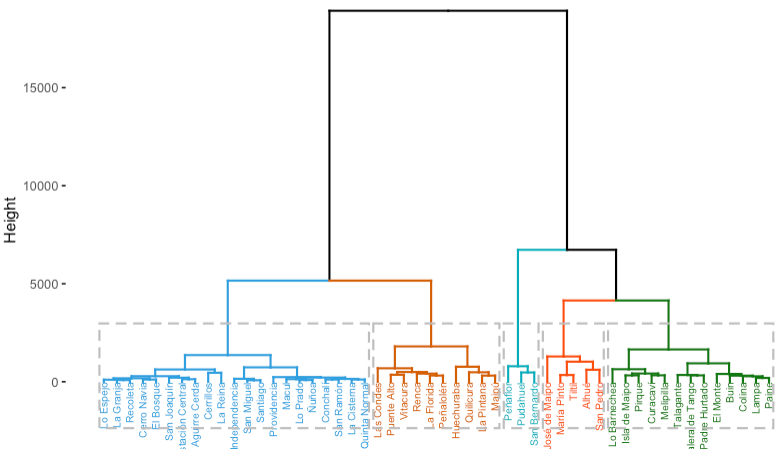}
	\caption{Hierarchical clustering using DTW for mobility time series. The solution with five clusters is highlighted.}
	\label{fig:2}
\end{figure}

The figure \ref{fig:2} shows a dendrogram, the result of applying a hierarchical analysis with DTW distance and complete linkage method, where the  5 clusters solution has been highlighted. In essence, the analysis recognizes similar behaviors of communes according to the value of the mobility Scores. The Cluster Validity Indices (CVI) are shown in the table \ref{table1}. The solution with 2 clusters has the best indices, however, fundamentally this solution indicates that there are communes with low mobility (rural communes) and communes with medium/high mobility (urban communes). We note that the mobility scores are not adjusted according to the population of the commune. We consider that for the purposes of the contagion analysis, crude or absolute score is a better indicator. Thus, for a more detailed analysis, the next best solution is the one corresponding to 5 clusters.

\begin{figure}[H]
	\centering
	\includegraphics[trim = 0cm 0cm 0cm 0cm, width = 1.0\textwidth]{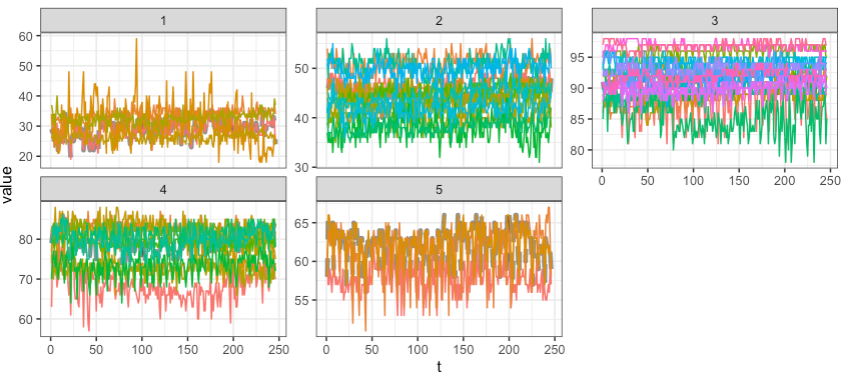}
	\caption{The different clusters and the time series that belong to those clusters. }
	\label{fig:3}
\end{figure}

\begin{figure}[H]
	\centering
	\includegraphics[trim = 0cm 0cm 0cm 0cm, width = 1.0\textwidth]{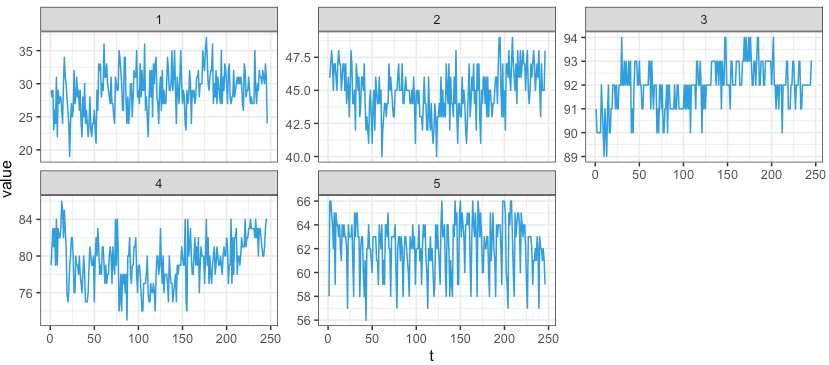}
	\caption{The lines represent the centroid time series for each cluster.}
	\label{fig:4}
\end{figure}

The figure \ref{fig:3} shows each of the clusters with the time series of their respective members.  It is observed that each cluster includes communes with close mobility scores. Thus, cluster 1 is made up of the communes with the lowest mobility Scores that belong to the approximate score interval (20 - 40). Cluster 2 (40 - 50), Cluster 3 (80 - 100) is made up of the communes with the highest mobility Scores. Cluster 4 (60 - 80) and cluster 5 (55 - 65) have some overlap. These behaviors are better appreciated in the figure \ref{fig:4}, which shows the time series of mobility Scores for the representative centroids of each cluster.

\begin{figure}[H]
 	\centering
 	\includegraphics[trim = 0cm 0cm 0cm 0cm, width = 1.0\textwidth]{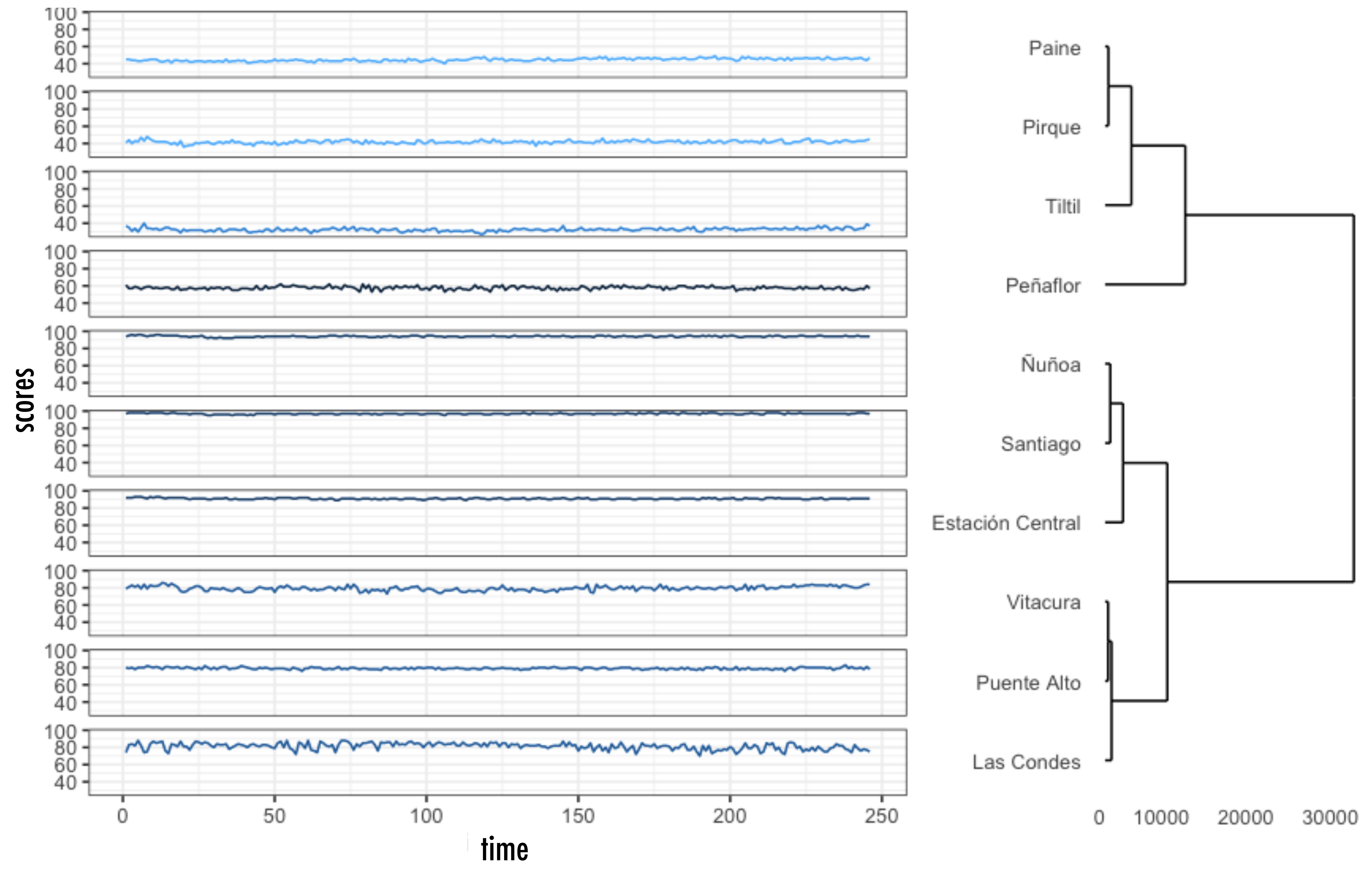}
 	\caption{A sample of 10 communes grouped in clusters of similar behavior according to the value of the scores. The communes of the center of the Metropolitan Region have the highest scores, while the rural communes of the periphery have the lowest scores. It is also observed that communes with different income levels may have similar mobility scores (Puente Alto, VItacura, Las Condes).}
 	\label{fig:5}
 \end{figure}

For a better visualization, the figure \ref{fig:5} shows a sample of 10 communes in the MR and their corresponding membership to a cluster according to their mobility Score values. The two-cluster solution clearly shows that the urban communes (Ñuñoa, Santiago, Estación Central, Vitacura, Puente Alto and Las Condes) have high mobility Scores compared to rural communes (Paine, Pirque, Tiltil and Peñaflor), which have less mobility Scores. However, the structure of the cluster shows some more interesting details that this graphic helps to highlight. For example, among the communes belonging to the cluster with low Scores, Peñaflor clearly stands out with more higher Scores.  This fact could be explained by the presence of several industries in the area that clearly did not stop operating during the months of the pandemic.
 
The central communes of the MR (Santiago, Ñuñoa and Estación Central) are the ones with highest mobility Scores. This can be explained fundamentally by the population density and/or the large number of services in the area. An observation of relevance for our subsequent analysis is given by the fact that communes with very dissimilar incomes such as Las Condes, Vitacura (two of the communes with the highest income in the country) and Puente Alto (a commune with very low income) have very similar mobility Scores. Puente Alto is the commune with more COVID-19 cases in the MR, as indicated by the analyzed COVID-19 dataset. However, the communes with the highest income, mentioned above, have significant fewer COVID-19 cases. This fact indicates that mobility is not the only factor to consider in the COVID-19 spread. Simply, community-specific infection risk also reflects others commune attributes that include income, education, health-care access, cultural practices, among other possible factors.

\begin{table}[!ht]
\begin{center}
\begin{tabular}{|c|c|c|c|c|c|c|c|}
	\hline
	Num. of Clusters& Sil &  SF &  CH & DB  & DBstar  & D  & COP \\
	\hline
	2 & 0.8 &  0.0& 158.6& 0.2 & 0.2 &0.1  & 0.1 \\
	\hline
	3& 0.6 & 0.0 & 94.0 & 0.2 & 0.3 & 0.1 & 0.1\\
	\hline
	4& 0.7 & 0.0 & 134.1 & 0.3 & 0.4 & 0.1 & 0.1 \\
	\hline
	5&  0.7& 0.0 & 140.6 & 0.3 & 0.3 & 0.1 & 0.0 \\
	\hline
	 6& 0.6& 0.0 & 128.5 & 0.3 & 0.4 & 0.1 &0.0  \\
	\hline
\end{tabular}
\end{center}
\caption{
	{\bf Cluster Validity Indices.}}
\begin{flushleft} \textbf{Sil}: Silhouette index \cite{Arbelaitz}; (to be maximized), 
	\textbf{SF}: Score Function \cite{Saitta}; (to be maximized),
	\textbf{CH}: Calinski-Harabasz index \cite{Arbelaitz}; (to be maximized),
	\textbf{DB}: Davies-Bouldin index \cite{Arbelaitz}; to be minimized),
	\textbf{DBstar}: Modified Davies-Bouldin index (DB*) \cite{Kim}; (to be minimized),
	\textbf{D}: Dunn index \cite{Arbelaitz}; (to be maximized),
	\textbf{COP}: COP index \cite{Arbelaitz}; (to be minimized).
\end{flushleft}
\label{table1}
\end{table}

\subsection{Mobility indices from the mobile phone antenna network.}

Another source of mobility data is the ubiquitous infrastructure provided by a cell phone network (see figure \ref{fig:1}B). These types of networks are built using a set of cell towers, which connect cell phones to the network.  Each cell tower has a latitude and a longitude - its geolocation -  and gives cellular coverage to an area called a sector. We assume that each sector is a 2-dimensional non-overlapping polygon, that can be geometrically modeled. The exact position of the cell phone user within each sector is unknown.

When a cellular device interacts with the network, there is a record of its connection to the antenna and a timestamp of when this occurred. The passage from one antenna to another is considered a trip. The number of trips between communes is estimated as the sum of the trips between antennas found in each commune. The Mobility Index \cite{ids} corresponds to how many trips (antenna transitions) were made within a commune or between communes. 

\begin{figure}[H]
	\centering
	\includegraphics[trim = 0cm 0cm 0cm 0cm, width = 1.04\textwidth]{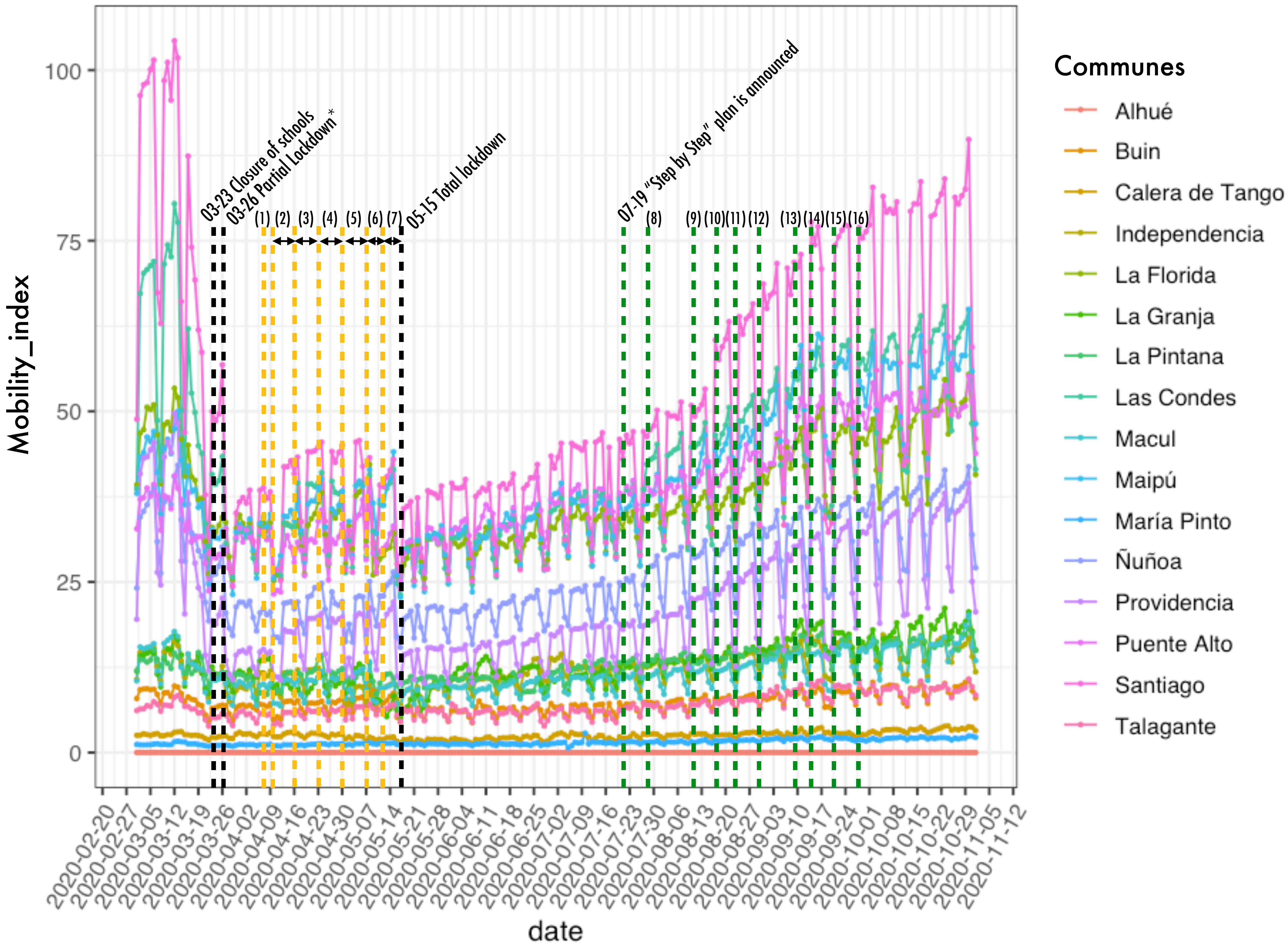}
	\caption{Mobility Index for a sample of 16 MR communes corresponding to the period from 2020-02-27 to 2020-11-05. The dates that correspond to the most important confinement measures are highlighted.}
	\label{fig:7}
\end{figure}

The figure  \ref{fig:7} shows the time series for the Mobility Index in the observation time period from 2020-02-27 to 2020-11-05 for a sample of 16 communes of the MR. The figure also includes the dates that correspond to the most important confinement measures taken in this time interval\footnote{The numbers in the figure indicate the dates with main confinement events. Details can be found in the Appendix A }. The Mobility Index, as shown in figure \ref{fig:7}, makes it possible to capture the impacts on mobility that were caused by the application of successive confinement measures. It is important to highlight the appreciable effect in the reduction of mobility produced by the closure of schools and later, to a much lesser extent, the different dynamic quarantines of the communes in the first days of the pandemic.

As of 2020-05-15, a total quarantine was declared for most communes in the MR. In the first days of implementation of this restriction measure, as can be seen in figure \ref{fig:7}, a slight decrease in mobility was recorded in all communes. However, mobility was increasing steadily until reaching the same levels as before the application of quarantine, showing signs of ``fatigue'' in complying with this restriction measure over time. With the gradual exit of the communes from quarantine, a very abrupt increase in mobility is observed.

\begin{figure}[H]
	\centering
	\includegraphics[trim = 0cm 0cm 0cm 0cm, width = 0.7\textwidth]{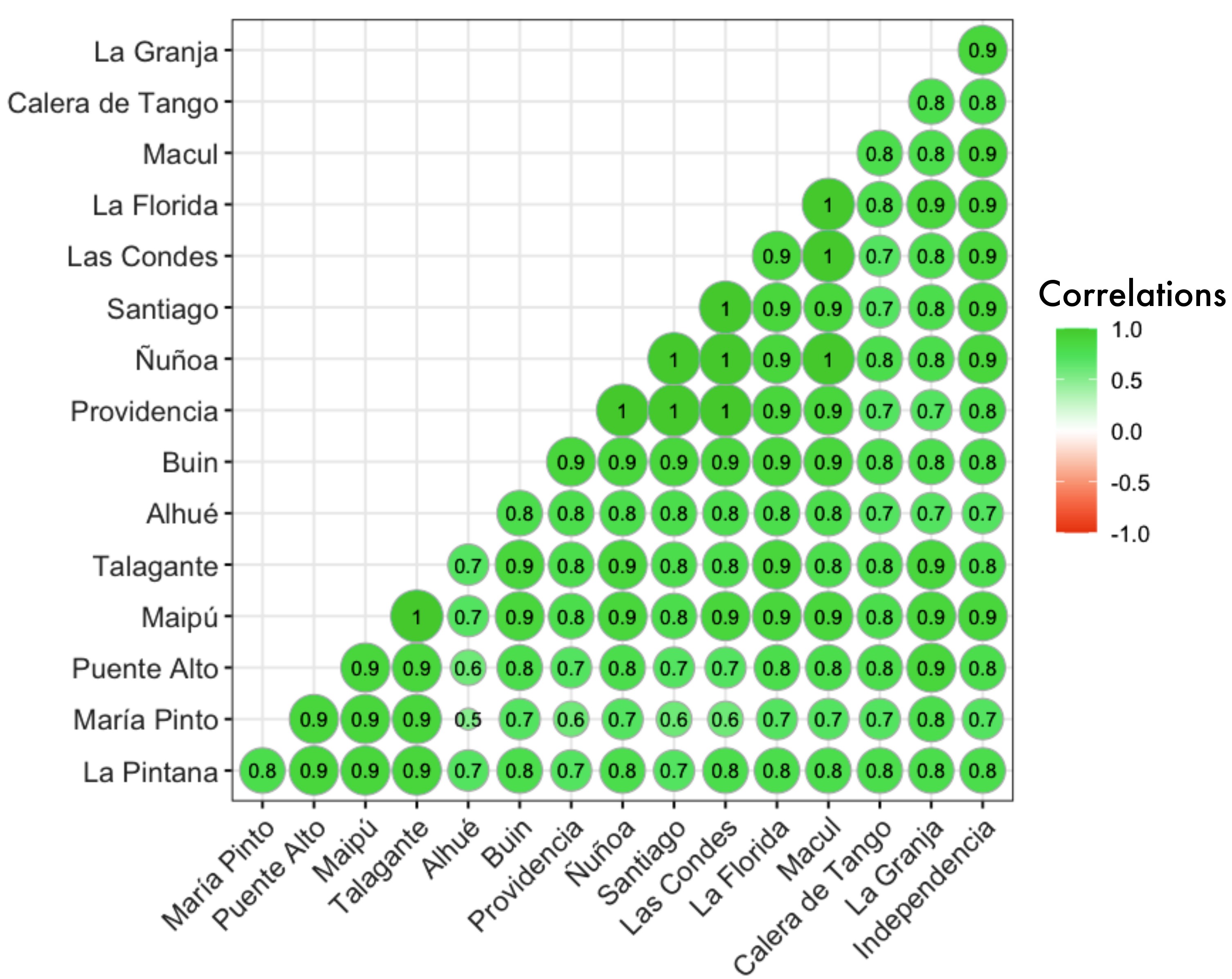}
	\caption{Mobility index correlations for MR communes}
	\label{fig:8}
\end{figure}

Another aspect to highlight is the strong correlation of the Mobility Index values between all communes of the MR as it is shown in figure \ref{fig:8}. This correlation in practice means that to a high degree the evolution of the mobility, measured by the Mobility Index, acquires a high level of synchronization, in the sense that, if in a commune the Mobility index rises/falls steadily, then this index rises/falls steadily in the rest of the other MR communes. The effect of this global behavior can also be seen in the analysis of the average Mobility Score (see figure \ref{fig:6}). The mobility Score describes the relative mobility (percentiles in the mobility distribution) and shows high regularity and stability throughout the observation time for each commune. This regularity is sustained over the time with few variations, and is not affected by confinement interventions. This fact, unlike the Mobility index,  makes the mobility Score of little use, specifically to capture the effect of the confinement policies in the mobility of the MR communes.

\section{Models}

\subsection{Urban public transport seeded the COVID-19 pandemic across the Santiago Metropolitan Region.} 

Pre-existing mobility patterns determine the routes along which diseases may potentially spread \cite{Hufnagel}.
Before the pandemic, like all cities, Santiago MR had its well-defined and regular mobility patterns. These patterns were remarkably well marked in the daily flow of users of the public transport network. 

\begin{figure}[H]
	\centering
	\includegraphics[trim = 0cm 0cm 0cm 0cm, width = 1.0\textwidth]{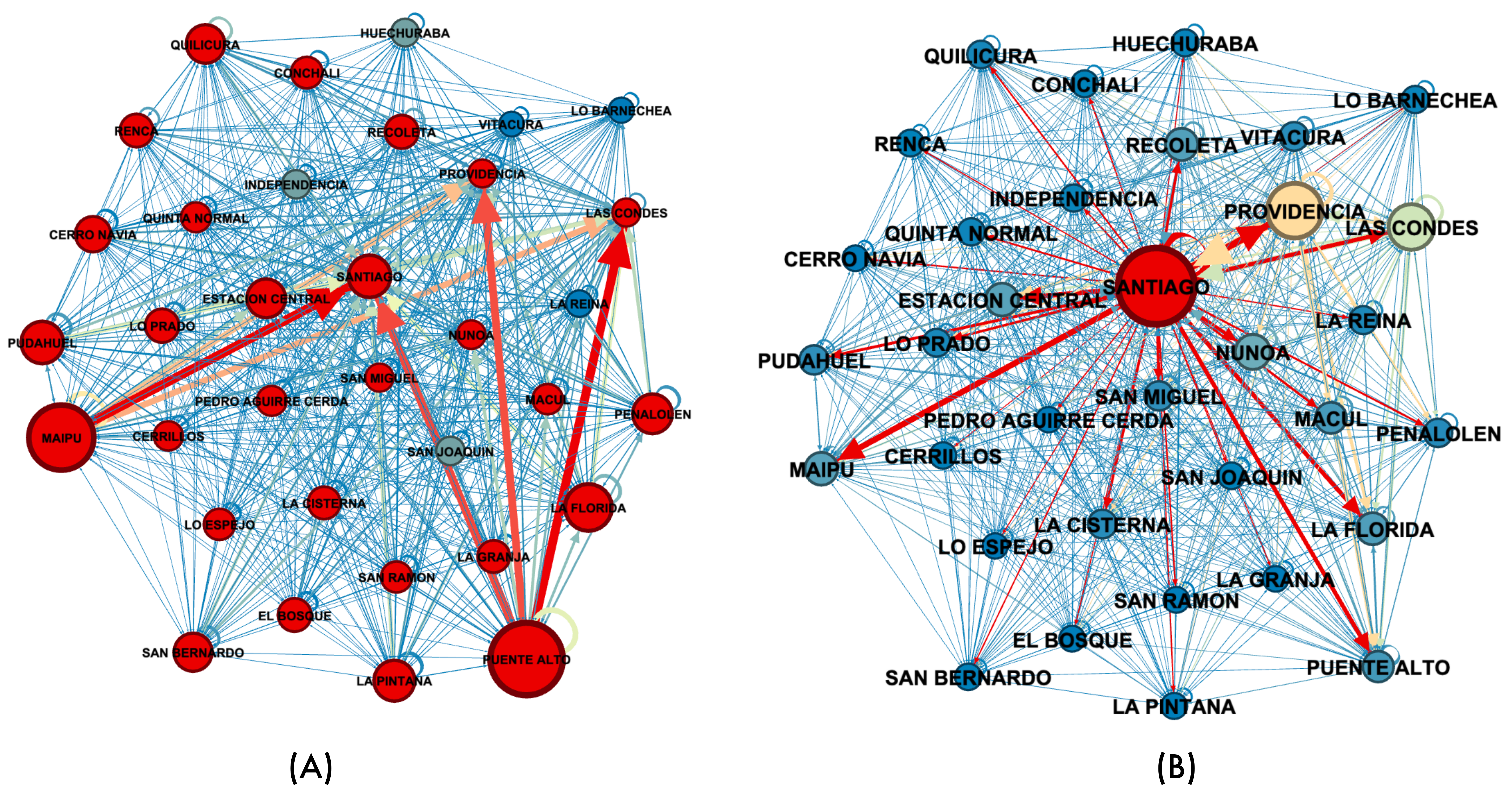}
	\caption{ Origin-destination between communes of Santiago City at 6:00 am (a) and 6:00 pm (b) in a regular pre-pandemic week day. The sizes of the arrows are related to the number of journeys, and the sizes of the nodes with its out-degrees.}
	\label{fig:11}
\end{figure}

In figure \ref{fig:11} the communes of origin and destination connected by the public transport service are shown at two specific times for a given working day of the week. At 6:00 AM (panel A), mass exits to workplaces occur, while at 6:00 PM (panel B) the return to homes begins. In the figure, the nodes represent the communes (only the urban ones are represented) and the links are the displacements between them. The thickness of the link between two communes is given in reference to its weight, that is, the number of people leaving the commune of departure towards the commune of destination, which is marked by the head of the arrow (directed link). The diameter of the node is proportional to the outdegree, which is measured considering the total weight of the links leaving the node. The pattern we want to highlight is the massive displacement from the most populated communes (which are also the ones with the lowest incomes) to the communes with the highest income at 6:00 AM.  For example, transportation from communes such as Puente Alto and Maipú to Las Condes and Providencia stands out. Likewise, the opposite effect is shown at 6:00 p.m., the hour at which the return to homes begins. 

 During the early days of the pandemic, these are the kinds of patterns that occurred regularly on the working days of the week\footnote{It is good to note that due to the so-called social outbreak prior to the pandemic, displacement patterns were affected. However, these events did not substantially change global trends related to displacement to workplaces. Even under adverse conditions, people have to go out to work to survive.}. 

The first cases of COVID-19 took place precisely in the richest communes of the MR. As of March 30, 2020, among the communes with the highest rate in the whole country, were Lo Barnechea 54.8 (68 cases), Las Condes 54.7 (181 cases) and Providencia 52.0 (82 cases), the three communes with the highest per capita income in the country. Taking into account the pattern of mobility in public transport, it could be inferred how the poorest communes became infected quickly. By the time the authorities made the decision to partially confine Puente Alto (2020-04-08) for example, the number of cases in this commune had risen to 239 (it should be noted that the number of PCRs taken in this commune was still low, so the number of actual cases was probably much higher). 

Exploratory data analysis suggests that the low-income communes of the MR with the highest regular flow to and from wealthy communes with a high prevalence of COVID-19 in the early stages of the pandemic, later experienced a greater growth in the number of cases. This, in turn, also suggests that urban public transport was a vector for the spread of contagion in the initial phase of the pandemic.

Let's take a closer look at these suggestions considering the early days of the pandemic. The highest number of infections occurred in the richest communes of the MR, while in the poorest the number of cases was relatively low. In this stage, the contagion rate is very high and mobility plays a preponderant role. Although the closure of educational establishments was already in force, this fact did not affect the displacements from the poorest to the richest communes, which are mainly due to work reasons. Therefore, we can hypothesize that the spread of contagion is mainly due to connectivity between communes, especially with those that we could call ``high-risk communes'' due to the high number of infections in them. Public transport thus becomes the main diffuser of the virus at this stage of the pandemic.

To quantitatively describe this situation, the flow of public transport users from low-income communes (with fewer cases) to richer communes with a higher contagion rate and a greater number of COVID-19 cases is considered. Without any other measure restricting mobility, the flows between communes  follow regular and well-established patterns, observed in the public transport of the MR on daily base (see figure \ref{fig:11}). So, we estimate the flow between communes by aggregating the number of trips registered with the origin-destination matrices \cite{bip}. We only consider trips between 6:00 AM and 8:00 AM, as these trips are mainly motivated by work reasons. In the late afternoon, at the end of the working day, mainly from 6:00 p.m., users return to the communes of origin, eventually carrying the virus with them.

To validate this hypothesis, which essentially states that during the first stage of the pandemic the virus spreads from the richest to the poorest communes, with a predominant role played by public transport, we built the following regression model:
\begin{equation}\label{OLS1}
\text{CumCases}_i =\beta_1\cdot \text{Flow}_i+ \beta_2\cdot \text{Flow}_i\times \text{MobOut}_i+\beta_3\cdot \text{MobOut}_i+ \beta_4\cdot \text{MobIn}_i+ \beta_5 \cdot \text{Score}_i+\beta_0  + \epsilon_i	
\end{equation}

Where, Flow$_i$ is the the sum of the trips between 6:00 AM and 8:00 AM from commune $i$ to high-risk communes normalized by the number of inhabitants of commune $i$. We consider the interval from February 20 to March 30. The variable Score$_i$ is the average score\footnote{The Score is defined in the second paragraph of the section \ref{datasets}} of commune $i$ in the given time interval. MobIn$_i$ and MobOut$_i$ are the average internal and external Mobility Indices\footnote{Mobility indices were described in the fourth paragraph of the section \ref{datasets}} respectively in the indicated interval for commune $i$, and CumCase$_i$ are the cumulative cases as of March 30 for commune $i$.

The fit of the regression model is shown in Appendix B. The $R^2 = 0.86$, but we also are interested in the $\beta_1$ coefficient, which measures the effect of flows on case growth. This value is positive, and as the statistical test indicates it is significantly different from zero. This result establishes that a high flow of connectivity between communes with a low number of cases and communes with a high number of cases through public transport produces a quantifiable growth of COVID-19 cases in the former. 

As we will see later in the next section this clear relationship that the model shows between the growth of cases and mobility (within and out of the commune) during the first days of the pandemic is no longer so clear in the subsequent stages. In fact, when confinement measures begin to operate, socioeconomic determinants become more noticeable and social inequities are more apparent.

\subsection{The role of confinement measures in the local evolution of the pandemic}

Mobility restrictions have been an important part of the response to the ongoing COVID-19 pandemic. They have been essential for preventing massive outbreaks and reducing systemic stress in hospitals and clinics. However, implementing these restrictions not only infringes on people's rights, it can also be cumbersome and expensive \cite{Cetron, Onishi}. Therefore, it is vitally important to understand the effectiveness of such measures in controlling the spread of contagion. 

Given this importance, numerous authors have addressed the issue of mobility restrictions both theoretically and empirically \cite{Colizza, Arino, Epstein, Bajardi} (to name a few). Most of these studies have focused on the role of international travel restrictions in the rate of spread and have shown that travel restrictions can reduce the rate at which a disease spreads from the source of infection. However, researches related to local propagation in communities, or through public transport in cities, are less common. The work of Espinoza B. et. al \cite{Espinoza}, for example, focuses on dimensions of epidemics other than contagion rate, such as duration or final size in individual communities, also taking into account the role of availability and quality of health care. They showed that in certain cases, restriction measures are counterproductive for low-income communities. 

In our study we show, along the same lines, based on the available data that the restrictions of mobility in effect could eventually have a significant reducing impact on the size of the pandemic in the country, but only if the high-risk communes, which in the case of the Santiago MR are the high-income communes,  are strictly and early isolated. This restrictive measure had to be taken once the virus reaches the country and establishes itself in high-income communities, where residents can afford international travels. However, the measure was taken late and became not only inefficient but also counterproductive for low-income communes with precarious health resources. In fact, the data suggest that mobility restrictions may have accelerated the contagion process and,  therefore led to a higher than expected number of cases within the mobility-regulated and low-income MR communes.

\begin{figure}[H]
	\centering
	\includegraphics[trim = 0cm 0cm 0cm 0cm, width = 1.0\textwidth]{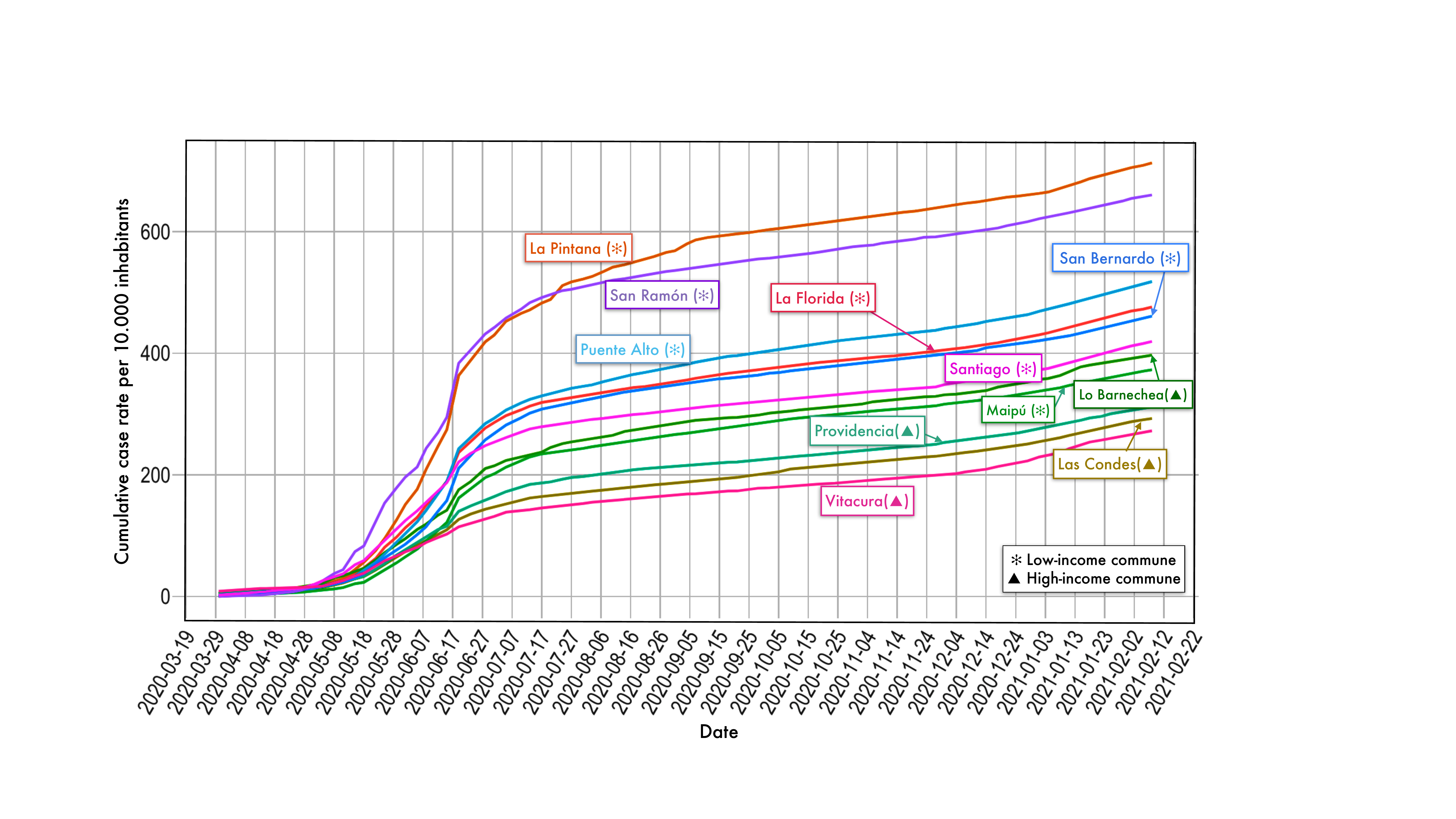}
	\caption{Cumulative cases of COVID-19 for the communes of the RM.}
	\label{fig:10}
\end{figure}

\begin{figure}[H]
	\centering
	\includegraphics[trim = 0cm 0cm 0cm 0cm, width = 1.0\textwidth]{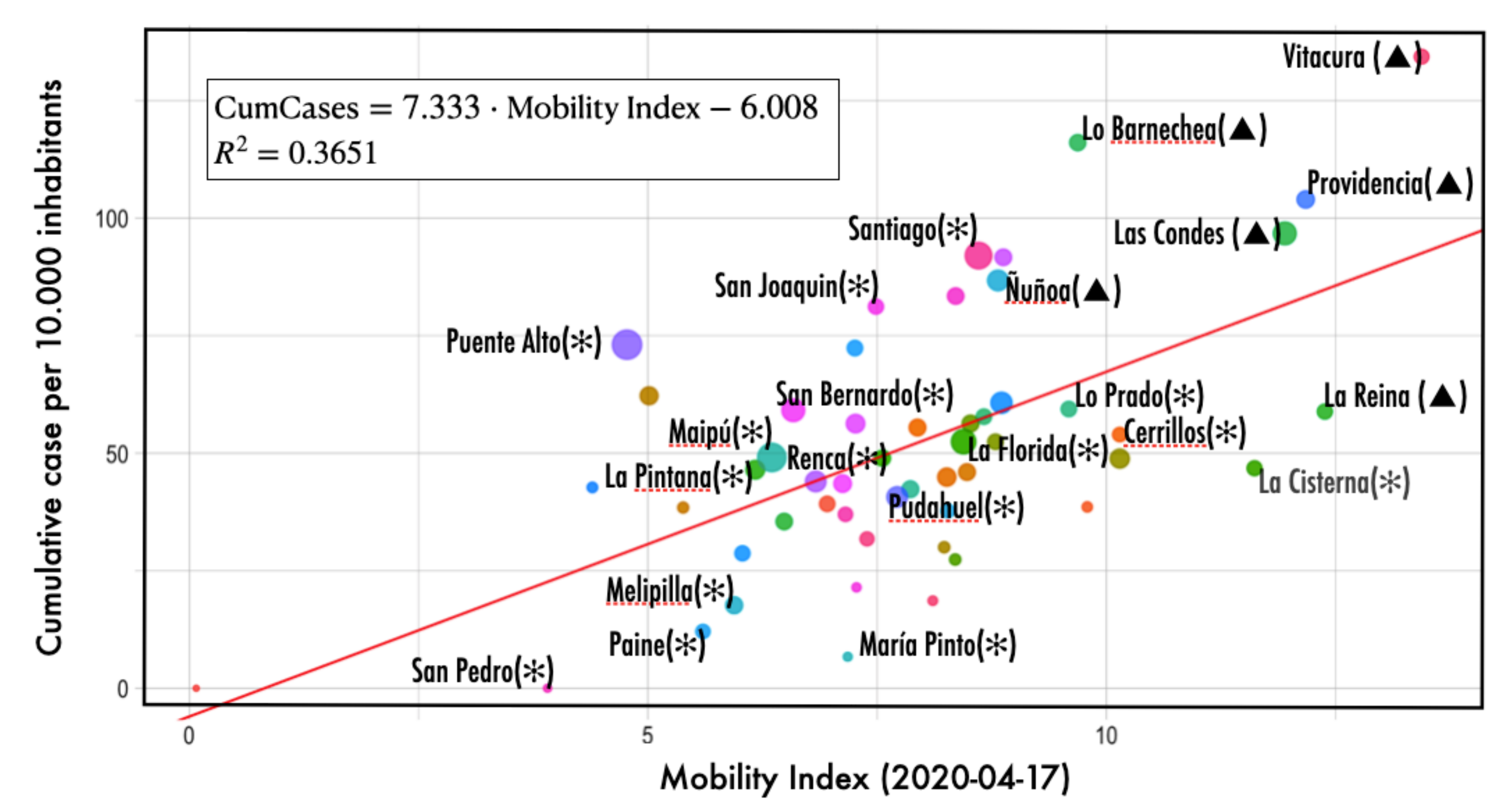}
	\caption{ Mobility index vs Cumulative number of cases per 100.000 inhabitants of the commune at the beginning of the total quarantine for the communes of the MR. The radius of the bubbles is proportional to the commune population. }
	\label{fig:14}
\end{figure}

\begin{figure}[H]
	\centering
	\includegraphics[trim = 0cm 0cm 0cm 0cm, width = 1.0\textwidth]{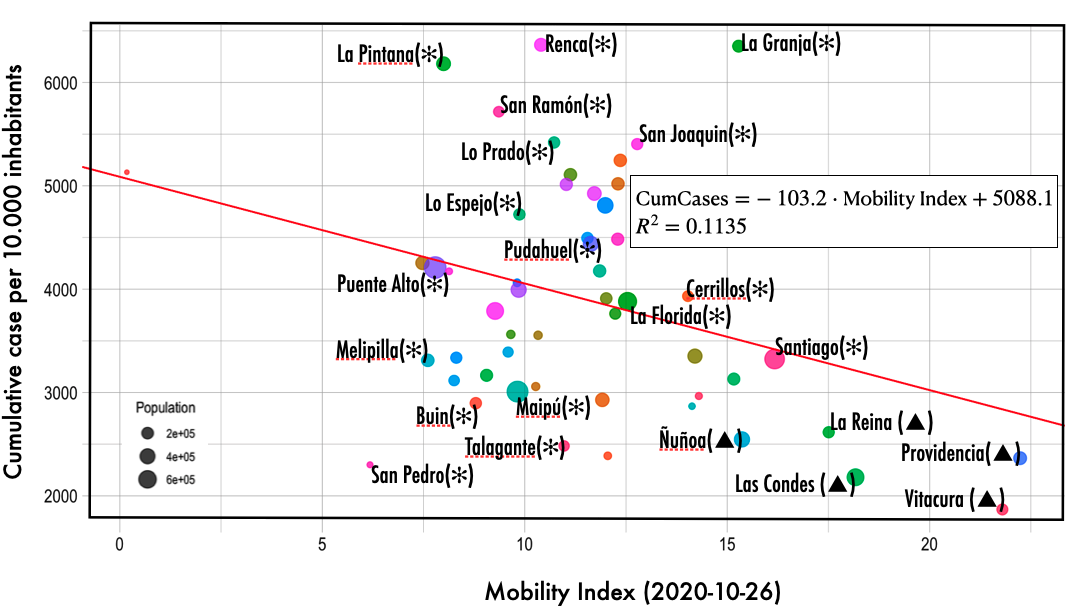}
	\caption{The repercussions left by measures that restricted mobility on the number of cases. The Mobility Index is shown versus the cumulative number of cases per 100.000 inhabitants after relaxing the containment measures.}
	\label{fig:15}
\end{figure}

It has been proposed that one of the most important factors in these adverse effects produced by the confinement measure is due to the fact that mobility restrictions are not effectively respected. For example, Bennett M. in \cite{Bennett} analyzed the effectiveness of lockdowns at the commune level in Chile during the first two months of the pandemic. She showed that lockdown measures were not effective in lower-income areas that were subject to quarantine at the same time that high-income communes. The cause of this, she explains, is because the mobility restrictions were not respected in the low-income communes. However, a more detailed and widely available analysis of mobility data shows that this is not the main reason. As the figure \ref{fig:7} shows that this is not so. During the first days of confinement, the effect of  restriction measures is appreciable in both poor and rich communes. Furthermore, as shown in figure \ref{fig:5}, the mobilities of poor communes such as Puente Alto and that of rich communes like Las Condes are very similar (in the latter being even higher as shown in the figures \ref{fig:14} and \ref{fig:15} in this section), but the number of cases is very different, disfavoring by far the poorest commune, Puente Alto. 

The figure \ref{fig:10} shows the cumulative number of cases per 100.000 inhabitants from March 19, 2020 to February 12, 2021 for a sample of Santiago MR communes. It is observed, a explosive growth of COVID-19 cases between the months of April and July. This type of growth is much more sustained and steeper for low-income communes (for example, La Pintana, San Ramón, Puente Alto). On the other hand, high-income communes (for example, Providencia, Vitacura, Las Condes) respond much better to lockdown measures and manage to lower contagion rates, ``flattening the curve'' more efficiently. 

\begin{figure}[H]
	\centering
	\includegraphics[trim = 0cm 0cm 0cm 0cm, width = 0.9\textwidth]{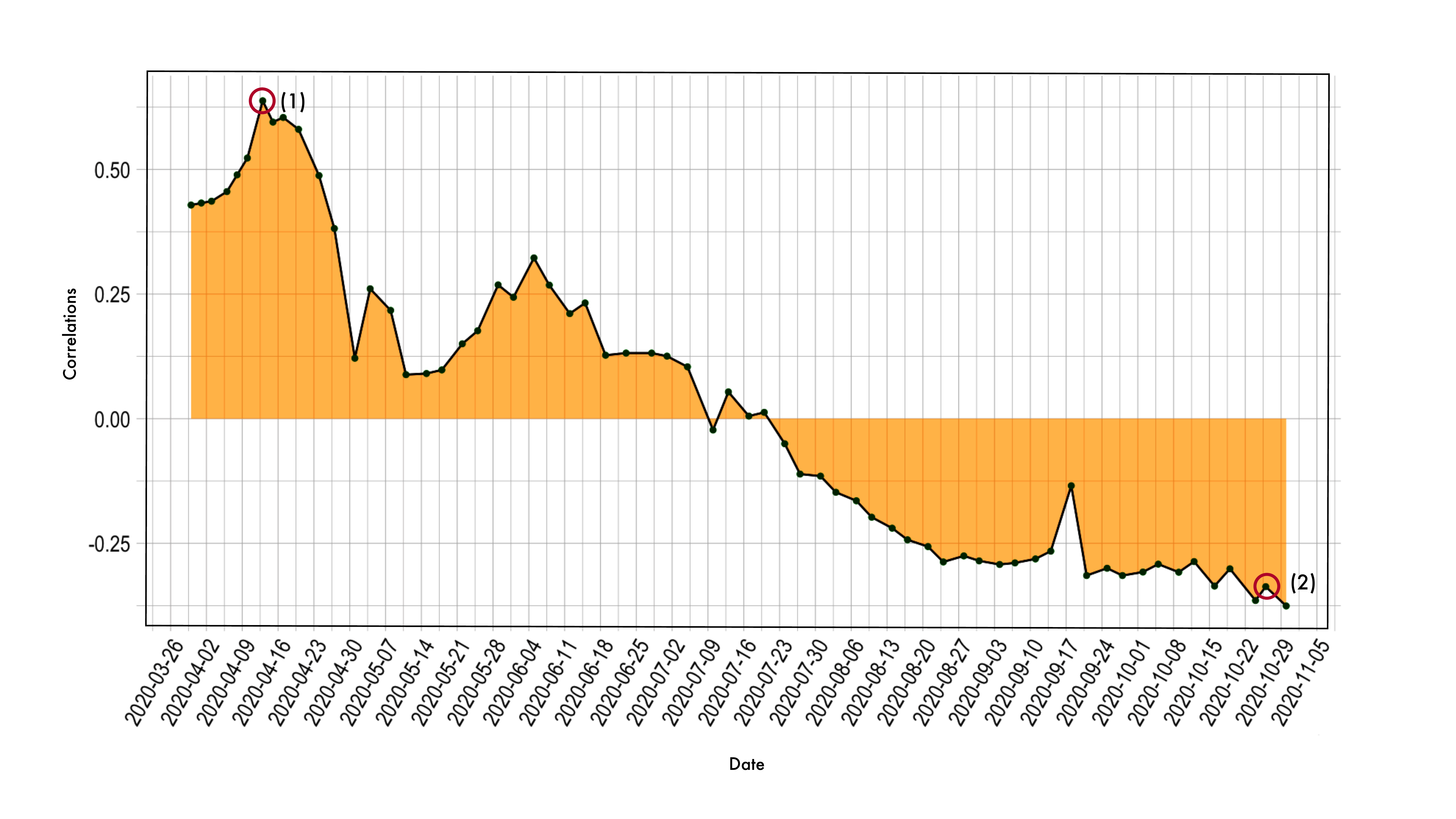}
	\caption{Temporal evolution of the correlation between the Mobility Index and the cumulative number of COVID-19 cases per 100.000 inhabitants in the period from March 26 to October 30, 2021.}
	\label{fig:13}
\end{figure}

The different socioeconomic conditions of the MR communes and the inequity in resources are manifested very clearly at this stage of the pandemic. Mobility stops playing an important role as it is shown by the correlation values between mobility and the number of COVID-19 cases in the figure \ref{fig:13}, which is practically null for this period.

In fact, in the first stage of the pandemic there is a strong positive correlation between mobility, measured by the Mobility Index, and the number of COVID-19 cases, but as of March 26, the partial and dynamic quarantines begin and this clear correlation begins to fade off.

The figure \ref{fig:13} shows the correlation between the Mobility Index and the cumulative number of cases per 100.000 inhabitants of the commune. This correlation is calculated for the period of time that goes from March 26 to October 30, 2020. Two dates in particular are marked in the figure, April 17 with a maximum positive correlation (0.6) and October 26 with a negative correlation (-0.34). 

On March 26, it is still observed that the communes with higher incomes and also with greater mobility have a greater number of cases (high-risk communes). Although other determinants should be considered, in addition to mobility, a regression model indicates that mobility alone explains about 37\% in the variability of the observed cases ($R^2\approx 0.37$) in this phase of the pandemic. The figure \ref{fig:14} shows a scatter diagram in which the horizontal axis stand for the Mobility Index and the vertical represents the cumulative number of COVID-19 cases per 100.000 inhabitants. The radius of the bubbles is proportional to the population in each commune. 

A characteristic pattern of this stage of the pandemic is observed. High-income communes have high mobility and a high number of cases, which makes them particularly active in spreading the virus. It could be inferred that a total and effective isolation of these communes in the early stages of the pandemic would have significantly reduced the magnitude of the contagion. On the other hand, more towards the center of the figure and above, with less mobility and with a smaller, but already significant number of cases, are the low-income communes. Further down and to the left are rural communes characterized by low mobility and a low number of cases.

As of May 15, the total quarantine comes into effect. However, by this date, the contagion has spread to the lowest-income communes and the measure seems to be taken late for these communes. Although there is still a positive correlation between mobility and the number of cases, this correlation is getting smaller and smaller. There is a slight decrease in mobility at the beginning of this period, but it never decreased significantly, as was shown in figure \ref{fig:7} that indicates a slow but sustained growth in mobility for this period. 

By October, the situation has changed. Mobility has been increasing and the communes with the highest incomes have the highest values in the Mobility Index, while the number of cases in these communes is decreasing. Figure \ref{fig:15} shows a scatter diagram illustrating this stage of the pandemic. Again, clear patterns can be seen in the distribution of the communes in the figure. On the right, below with high mobility and a low number of cases are the communes with the highest income. In the center, with less mobility but with a much higher number of cases, are the low-income communes. Further to the left and below, preserving a situation similar to that shown in figure \ref{fig:14} the rural communes are located. In this period, mobility is a variable that explains only 11\% of the variance of the observed cases ($R^2\approx 0.11$) and obviously other determinants, such as socioeconomic conditions and vulnerability in general, are fundamental to explain the observed contagion patterns.

The sustained increase in mobility and the presence still of a significant number of cases are the ingredients that trigger a second regrowth or ``second wave''. The first signs of this outbreak can already be seen in the figure \ref {fig:10}, where the curves of cumulative cases from the end of December begin to show an increase in the growth rates of COVID-19 cases.

\subsection{Assessing of the timely lifting of quarantine measures through Augmented Synthetic Control Method.}

Another relevant aspect of the analysis of the role of mobility restriction measures is precisely the timely lifting of these measures, especially quarantine. Since May 15, 2020 a long period of quarantine has been implemented in the MR. Although the measure was taken for practically all the communes of the MR simultaneously, the deconfinement occurred gradually. The communes were coming out of quarantine as the number of cases decreased. The first to leave were Colina, La Reina, Las Condes, Lo Barnechea, Til Til, Vitacura (basically some rural communes, along with the communes with the highest income). On the other hand, the communes with the longest confinement time are at a partial level: Puente Alto, specifically the western sector, which was under the measure for 172 days; and in their territorial totality: El Bosque and Quinta Normal, whose quarantines were in force for 158 days (all these communes are low-income). 

The case of Puente Alto is particularly interesting, given that it is the commune with the most COVID-19 cases in the country and where the lockdown measures were, therefore, inefficient. 

For the study of this aspect we used Synthetic Control \cite{Abadie}, in order to estimate the effect that the lifting of the quarantine had on the number of cases and also on the mobility of the affected communes.

\begin{figure}[H]
	\centering
	\includegraphics[trim = 0cm 0cm 0cm 0cm, width = 0.9\textwidth]{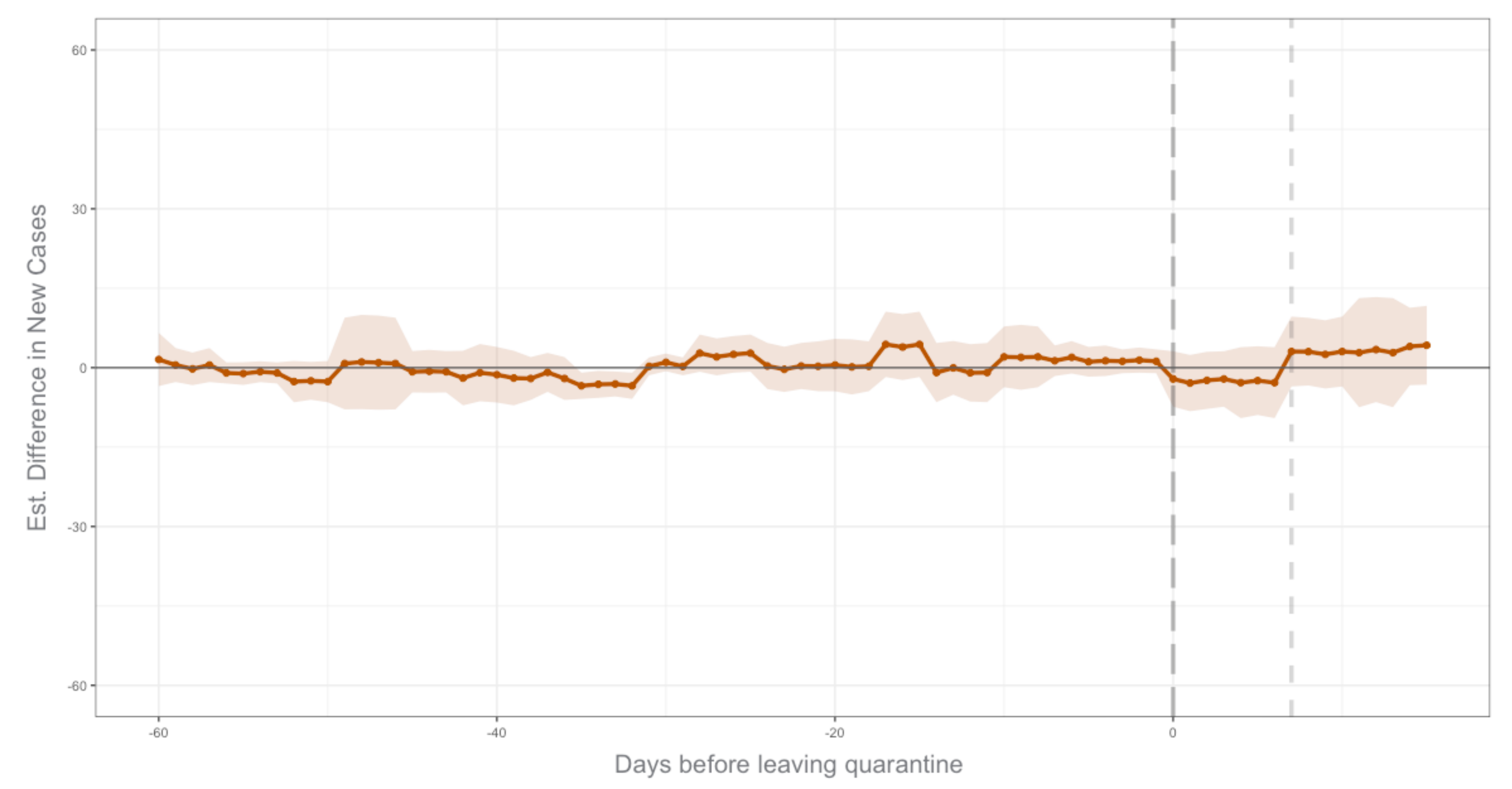}
	\caption{ Estimation of the effect of lifting the quarantine on COVID-19 new cases for  low-income communes in the MR. }
	\label{fig:12}
\end{figure}

For the analysis, we consider that the communes in the ``pre-treatment'' stage are all those that are in quarantine (practically all the communes of the MR for the observation time interval considered).

In this case, the ``treatment'' is lifting the quarantine. The communes were gradually leaving them. To consider this staggered lifting of the confinement measures and to be able to estimate the ``treatment effects'' we used the Augmented Synthetic Control Method (ASCM) \cite{Abadie, Ben-Michael}. The objective here is to estimate its consequences on mobility and on the number of COVID-19 cases for these communes by comparing with the opposite option, that is, having spent more days in quarantine.

The idea is to construct a weighted average of donor communes, known as a synthetic control, that matches the pre-treatment results of the treated communes. The estimated effect is then the difference in post-treatment outcomes between the treated communes and the synthetic control. The donors for the estimation were determined from the available group of communes and their memberships to different clusters previously defined from the similarities in mobility and also considering the socioeconomic conditions of the communes.

\begin{figure}[H]
	\centering
	\includegraphics[trim = 0cm 0cm 0cm 0cm, width = 0.9\textwidth]{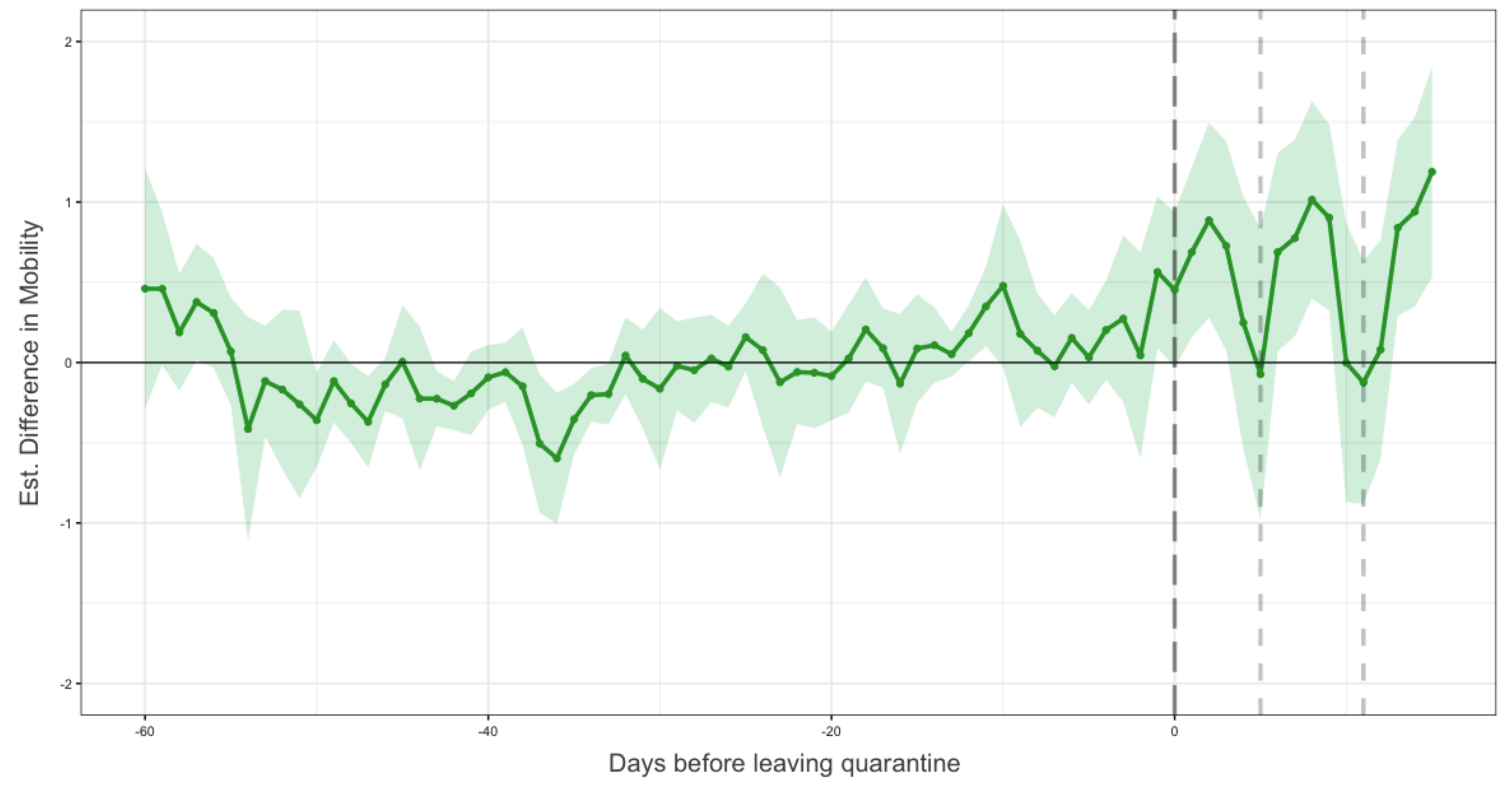}
	\caption{ Estimation of the effect of lifting the quarantine on Mobility in low-income communes of the MR. }
	\label{fig:16}
\end{figure}

The figure \ref{fig:12} shows the estimate of the effect of lifting the quarantine in the Recoleta, San Ramón, La Cisterna, La Granja, San Joaquín and San Miguel communes (all of them with low income) compared to the synthetic control built from other communes with similar socioeconomic characteristics that subsequently abandoned the quarantine (Quinta Normal, La Pintana, Lo Prado, Cerro Navia, Conchalí, Puente Alto and Lo Espejo). The figure describes how initially the number of COVID-19 cases, on average, for the communes treated tends to fall after lifting the restriction measure. However, this number is rapidly growing again, indicating that the measure may have caused a resurgence of contagion in the treated communes.

On the other hand, the figure \ref{fig:16} shows the gradual increase in mobility after lifting the confinement. However, since the uprising is not total, but the communes went to the so-called Phase 2 of the Step by Step plan, that is, the quarantine measure is maintained only on weekends, the periodicity in the estimate is clearly observed of mobility. On weekends, the difference with the mobility of the synthetic control decreases as once again the communes are temporarily quarantined.

\section{Discussion}

One of the objectives of this work is to make a consistent and validated narrative with the available data on the contagion patterns observed during the evolution of the pandemic in the MR of Santiago. This description of the events would help to understand how several interacting causes shape the observed patterns of virus spread and would support decision-making in future contagion developments. 

 The available data comprise 1) mobility data 2) socio-economic data and 3) COVID-19 case data.
The methodology used was based on 1) discovering/analyzing mobility patterns, 2) generating synthetic data to estimate causal relationships in the number of new cases of COVID-19 and Mobility after lifting the mobility restriction measures, 3) Analysis of correlations and regression models.

The initial contagion in the MR of Santiago begins in the richest communes where its inhabitants can afford international travels. Therefore, at the beginning of the pandemic, these communes are at high risk of contagion. An effective containment measure at this stage (creating a sanitary cordon around high-risk communes) could have reduced the size of the subsequent contagion across the country. This measure was not taken in time and the virus followed established routes of human mobility to spread. This occurs mainly through the public transport service. In fact, workers who have their source of work in high-risk communes were infected and carried the virus with them to their communes of origin. The effect is a rapid spread in both rich and low/middle-income communes, reinforced by the fact that workers use public transport to travel between communes. 

At this stage, there is a strong correlation between mobility and the number of cumulative cases of COVID-19 in each commune. Then the government begins with the policies of confinement and quarantines. The measure has a positive effect in the richest communes, which have greater resources and a better socioeconomic situation, but in the low/middle-income communes its effectiveness is poor, and there is even the cases of an acceleration in the number of infected for these communes. In most cases, far from a decrease, the reverse effect is observed.

There is no appreciable decrease in the internal mobility of the communes, but this happens in a similar way for both poor and rich communes. In low-income communes, it is impossible to stay home due to the lack of food and financial resources to survive in combination with little and late government aid. The effects are aggravated by the reduction in the number of services (in addition to its high geographic concentration that produces long lines and crowds of people) and the confinement measures prevent people from ``escaping'' away from the high-incidence places.

As a result, the contagion and the number of deaths grown in low-income communes, despite quarantines and taken measures. All this situation is aggravated by poor hygienic conditions, frequent base diseases in the population, poor diet, among other factors that further promote the spread and mortality of the virus. 

After a long time with the conditions described above, that is, high rates of infection and death, finally, the confinement measures begin to work and the number of cases begins to decrease. From here, the government begins to gradually relax the measures of restriction of mobility and gradually several communes leave the quarantine. However, the virus still remains in the population, mainly in some ``pockets'' located in low-income communes. With the explosive increase in mobility and the return to the old routes of movement, the conditions are again in place for the virus to spread again, thus beginning what some have called the ``second wave''. From here on, the government again takes up lock-down policies, this time through the ``Step by Step'' plan. However, restrictive measures lose many of their effectiveness. The long period lived with restrictions gives way to clear elements of ``fatigue''. On the one hand, people lose respect for these measures and on the other they simply do not have the economic resources to comply with them. Eventually, this creates the conditions for this second wave to have greater proportions or for a high prevalence level to be maintained for a long period of time, in which the number of cases cannot be reduced.

\pagebreak

\section*{Appendix A} 

\subsubsection*{Timeline of major events related to confinement measures for communes in the Metropolitan Region of Santiago from 2020/03/26 to 2020/09/28.}
\begin{verbatim} 
(*) 03/26 - 02/04 Las Condes, Vitacura, Lo Barnechea, Providencia, Santiago,
                  Ñuñoa, Independencia enter into partial lockdown.
(1) 02/04 - 08/04 Independence comes out of quarantine. 
(2) 04/09 - 04/16 Las Condes, Santiago (North), Ñuñoa (North), remain in 
                  quarantine. Puente Alto (West) enter into partial lockdown.
(3) 04/16 - 04/23 Only Santiago (North), Ñuñoa (North), El Bosque, 
                  San Bernardo (Northeast) remain in quarantine.
(4) 04/23 - 04/30 Santiago (North), Ñuñoa (North), Quinta Normal, PAC, 
                  El Bosque, San Bernardo (Northeast), Puente Alto (West) 
                  in partial lockdown.
(5) 04/30 - 05/07 Santiago (North), Ñuñoa (North), Quinta Normal, 
                  Estación Central,  PAC, Independencia, El Bosque, 
                  San Bernardo (Northeast), Puente Alto (West), San Ramón (South)
                  in partial lockdown.
(6) 05/07 - 05/12 Santiago, Quilicura, Recoleta, Independencia, Quinta Normal, 
                  Ñuñoa (North), Estación Central, Cerrillos, PAC, San Ramón (South), 
                  San Bernardo (Northeast), Puente Alto (West) in partial lockdown.
(7) 05/12 - 05/15 Total Lock-down for:  Santiago, Quilicura, Conchalí, Cerro Navia,
                  Renca, Recoleta, Independencia, Quinta Normal, Macul, Lo Espejo,
                  San Miguel, San Joaquín, Peñalolén, La Florida, La Granja, 
                  La Cisterna, La Pintana, Estación Central, Cerrillos, PAC, 
                  San Ramón, San Bernardo (Northeast),  Puente Alto (West)
(8) 07/28         The first communes go to phase 2, transition \cite{paso-paso}: 
                  Colina, La Reina, Las Condes, Lo Barnechea, Til Til, Vitacura. 
(9) 08 /10        Lampa, Melipilla y Providencia go to phase 2, transition.
(10) 08/17        Santiago y Estación Central go to phase 2, transition.
(11) 08/24        San José de Maipo, Peñalolén, Padre Hurtado y Peñaflor 
                  go to phase 2, transition. Paine enters quarantine.
(12) 08/31        La Florida, Maipú, Cerrillos, Calera de Tango, El Monte, PAC,
                  Macul, Talagante y Huechuraba go to phase 2, transition.
(13) 09/07        Recoleta, San Ramón, La Cisterna, La Granja, San Joaquín
                  y San Miguel go to phase 2, transition.
(14) 09/14        Quilicura, Isla de Maipo, San Bernardo go to phase 2, transition.
(15) 09/21        Pudahuel, Independencia y El Bosque go to phase 2, transition
(16) 09/28        Quinta Normal, La Pintana, Lo Prado, Cerro Navia, Buin, Conchalí, 
                  Puente Alto, Lo Espejo go to phase 2, transition.
\end{verbatim}

For an explanation of the meaning of phase 2 transition, see  \cite{paso-paso}.

\pagebreak

\section*{Appendix B} 

\begin{verbatim} 
	Coefficients:
                            Estimate Std. Error t value Pr(>|t|)    
(Intercept)                 -175.865     43.281  -4.063 0.000354 ***
MobIn                         22.627      2.272   9.957 1.06e-10 ***
MobOut                         6.989      6.385   1.095 0.283061    
Flow                          51.454     12.228   4.208 0.000240 ***
Score                          1.210      0.483   2.506 0.018292 *  
MobOut:Flow                   -7.276      1.738  -4.186 0.000255 ***
---
Signif. codes:  0 '***' 0.001 '**' 0.01 '*' 0.05 '.' 0.1 ' ' 1

Residual standard error: 25.67 on 28 degrees of freedom
Multiple R-squared:  0.8574,	Adjusted R-squared:  0.832 
F-statistic: 33.68 on 5 and 28 DF,  p-value: 5.178e-11
\end{verbatim}


\begin{thebibliography}{99}
	
\bibitem{grandata} URL https://covid.grandata.com

\bibitem{paso-paso} URL https://www.gob.cl/coronavirus/pasoapaso/

\bibitem{hex} URL https://h3geo.org/docs

\bibitem{movistar} URL https://github.com/MinCiencia/Datos-COVID19/tree/master/output/producto33

\bibitem{uai} Centro de Inteligencia Territorial 2012, Universidad Adolfo Ibañez, Santiago 2012

\bibitem{minciencia} URL https://github.com/MinCiencia/Datos-COVID19/blob/master/output/producto1/Covid-19.csv


\bibitem{dtw} Aghabozorgi, Saeed, Ali Seyed Shirkhorshidi, and Teh Ying Wah. (2015). ``Time-Series Clustering?A Decade Review.'' Information Systems 53. Elsevier: 16?38.


\bibitem{dtwR} Giorgino T, Tormene P (2009). dtw: Dynamic Time Warping Algorithms. R package version 1.13-1, URL http://CRAN.R-project.org/package=dtw.

\bibitem{Arbelaitz} Arbelaitz, O., Gurrutxaga, I., Muguerza, J., Perez, J. M., \& Perona, I. (2013). An extensive comparative study of cluster validity indices. Pattern Recognition, 46(1), 243-256.

\bibitem{Saitta} Saitta, S., Raphael, B., \& Smith, I. F. (2007). A bounded index for cluster validity. In International Workshop on Machine Learning and Data Mining in Pattern Recognition (pp. 174-187). Springer Berlin Heidelberg.

\bibitem{Kim} Kim, M., \& Ramakrishna, R. S. (2005). New indices for cluster validity assessment. Pattern Recognition Letters, 26(15), 2353-2363.
\bibitem{antenas}  http://datos.gob.cl/dataset/2019. Set of authorized, in Santiago, cell-phone antennas

\bibitem{ids} Universidad del Desarrollo (2020). Indice de Mobilidad Pandemia COVID-19. Technical Report Instituto Data Science. Universidad del Desarrollo and Telefonica.

\bibitem{bip} http://www.dtpm.gob.cl/index.php/documentos/matrices-de-viaje


\bibitem{minsalud1} Ministerio de Salud. MINSAL. Decimoséptimo informe epidemiológico enfermedad por COVID-19. Departamento de Epidemiología. Available at https://www.minsal.cl/wp-content/uploads/2020/05/Informe\_EPI\_15-05-20.pdf 


\bibitem{who1} World Health Organization (2020) COVID-19 Strategic Preparedness and Response Plan: Country Preparedness and Response Status for COVID-19. Geneva: WHO; June 09, 2020. Available at https://www.who.int/publications/i/item/updated-country-preparedness-and-response-status-for-covid-19-as-of-9-june-2020

\bibitem{lowy} https://interactives.lowyinstitute.org/features/covid-performance/

\bibitem{complex1} New England Complex Systems Institute. Countries beating COVID-19. EndCoronavirus.org 2021. Available at https://www.endcoronavirus.org/ countries\#winning (Accessed 3 February 2021). 

\bibitem{Hufnagel} Hufnagel L, Brockmann D, Geisel T. Forecast and control of epidemics in a globalized world. Proceedings of the National Academy of Sciences. 2004;101(42):15124?15129.


\bibitem{Cetron} Cetron M, Landwirth J. Public health and ethical considerations in planning for quarantine. The Yale journal of biology and medicine. 2005;78(5):329. pmid:17132339


\bibitem{Onishi} Onishi N. As Ebola Grips Liberia?s Capital, a Quarantine Sows Social Chaos; 2014. Available from: https://www.nytimes.com/2014/08/29/world/africa/in-liberias-capital-an-ebola-outbreak-like-no-other.html [cited March 14, 2018].

\bibitem{Espinoza} Espinoza B, Castillo-Chavez C, Perrings C (2020) Mobility restrictions for the control of epidemics: When do they work? PLoS ONE 15(7): e0235731. https://doi.org/10.1371/journal.pone.0235731


\bibitem{Colizza} Colizza V, Vespignani A. Epidemic modeling in metapopulation systems with heterogeneous coupling pattern: Theory and simulations. Journal of theoretical biology. 2008;251(3):450?467. pmid:18222487

\bibitem{Arino} Arino J, Jordan R, Van den Driessche P. Quarantine in a multi-species epidemic model with spatial dynamics. Mathematical biosciences. 2007;206(1):46?60. pmid:16343557

\bibitem{Epstein} Epstein JM, Goedecke DM, Yu F, Morris RJ, Wagener DK, Bobashev GV. Controlling pandemic flu: the value of international air travel restrictions. PloS one. 2007;2(5):e401. pmid:17476323

\bibitem{Bajardi} Bajardi P, Poletto C, Ramasco JJ, Tizzoni M, Colizza V, Vespignani A. Human mobility networks, travel restrictions, and the global spread of 2009 H1N1 pandemic. PloS one. 2011;6(1):e16591. pmid:21304943

\bibitem{Bennett} Bennett M,. All things equal? Heterogeneity in policy effectiveness against COVID-19 spread in chile, World Development, Vol. 137, 2021, https://doi.org/10.1016/j.worlddev.2020.105208.

\bibitem{Abadie} Abadie, A., Diamond, A., \& Hainmueller, J. (2015). Comparative politics and the synthetic control method. American Journal of Political Science, 59, 495?510.

\bibitem{Ben-Michael} Ben-Michael, E., Feller, A., \& Rothstein, J. (2020). The augmented synthetic control method. Working Paper, UC Berkeley



\end{thebibliography}
\end{document}